\documentclass[]{spie}  %>>> use for US letter paper
\usepackage{amsmath,amsfonts,amssymb}
\usepackage{graphicx}
\usepackage[colorlinks=true, allcolors=blue]{hyperref}
\usepackage{booktabs}

\title{Optimizing photon-number resolution with superconducting nanowire multi-photon detectors}

\author[1,2]{Timon Schapeler}
\author[3]{Fabian Schlue}
\author[3]{Michael Stefszky}
\author[3]{Benjamin Brecht}
\author[3]{Christine Silberhorn}
\author[1,2]{Tim J. Bartley}
\affil[1]{Department of Physics, Paderborn University, Warburger Str. 100, 33098 Paderborn, Germany}
\affil[2]{Institute for Photonic Quantum Systems (PhoQS), Paderborn University, Warburger Str. 100, 33098 Paderborn, Germany}
\affil[3]{Integrated Quantum Optics Group, Institute for Photonic Quantum Systems (PhoQS), Paderborn University, Warburger Str. 100, 33098 Paderborn, Germany}

\authorinfo{Further author information: Send correspondence to T.S., E-mail: timon.schapeler@upb.de}

\begin{document} 
\maketitle

\begin{abstract}
In this paper we briefly review the state-of-the-art of intrinsic photon-number resolution (PNR) with superconducting nanowire single-photon detectors (SNSPDs) and will provide an overview of the various experimental techniques employed to achieve PNR. Additionally, we introduce a resolvability criterion to PNR based on SNSPDs.
\end{abstract}

\keywords{superconducting nanowire single-photon detector (SNSPDs), photon-number resolution (PNR), photon counting, review, resolvability criterion}

\section{Introduction} \label{sec:intro}
The ability to precisely determine the number of photons in a light field is a crucial capability for a wide range of quantum optics experiments and applications, including quantum communication~\cite{gisin2002quantum,gisin2007quantum}, quantum computing~\cite{kok2007linear}, and metrology~\cite{slussarenko2017unconditional}.
%Superconducting nanowire single-photon detectors (SNSPDs) have long been regarded as a threshold detectors, i.e., their response to any non-zero number of incident photons was regarded as a ``click'' event and if no photon are incident as a ``no-click" event. However, recent investigations have shown that a careful analysis of the electrical output signal of the detector can give limited photon-number information.
While counting photons is a relatively simple task conceptually, it is surprisingly difficult to quantify the ``photon-number resolvability'' of a given detector. In other words, there is no consensus on how to answer the question: how many photons can your detector count? For highly sensitive bolometers such as transition edge sensors (TESs)~\cite{cabrera1998detection,lita2008counting}, one can simply measure the energy deposited, and if a monochromatic light source is used, the number of absorbed photons can be inferred. But even these devices have a finite energy resolution, which leads to uncertainties in the inferred photon number. In essence, this difficulty arises from the discrete nature of photons, but continuous nature of measurement outcome distributions. 

In this paper, we first review the state-of-the-art of intrinsic PNR with SNSPDs. We then propose a resolvability criterion for photon-number measurement outcome distributions from SNSPDs, to determine the highest number of photons that can be resolved with an SNSPD in a certain configuration.
We apply this measure to our own measurements of the intrinsic photon-number resolution based on the arrival time of output pulses from an SNSPD. Finally, we give an outlook on the remaining challenges for PNR with SNSPDs.

\section{State of the art}\label{sec:stateoftheart}
Known for their high detection efficiency~\cite{reddy2020superconducting,chang2021detecting}, low timing jitter~\cite{korzh2020demonstration}, fast reset times~\cite{muenzberg2018superconducting} and low dark count rates~\cite{chiles2022new}, SNSPDs have primarily been utilized as binary detectors, capable of only distinguishing between the absence and presence of one or more photons. Nevertheless, in a certain regime of measuring short, few-photon pulses, some intrinsic PNR can be obtained. 

\begin{figure}[t]
    \centering
    \includegraphics[width=1\linewidth]{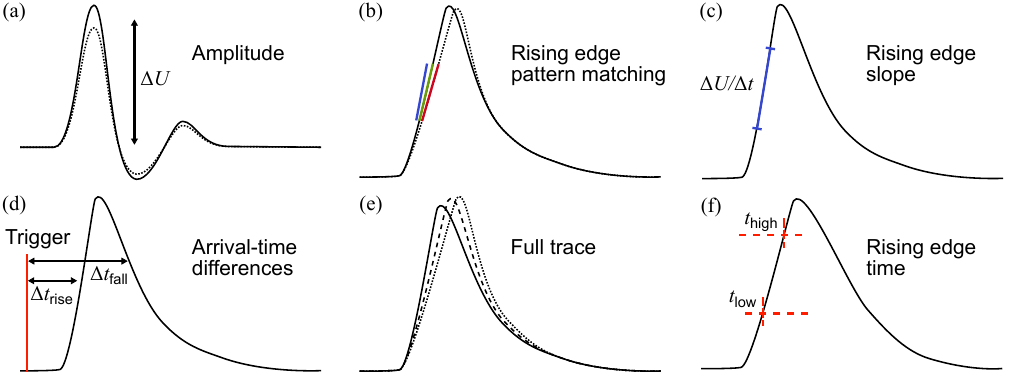}
    \caption{Schematic representation of the different techniques to retrieve photon-number information from the electrical output signals of SNSPDs. (a) The amplitude of the signal is sensitive to the number of photons, either through an inductor-resistor differentiating circuit or an impedance-matching taper. (b) Different colored lines represent reference waveforms of a subset of the rising edge. (c) A linear fit is made to the rising edge slope. (d) Arrival-time differences from a trigger signal to the rising edge or both the rising and falling edge are used to retrieve photon-number information. This can be done with an oscilloscope or a time-to-digital converter. (e) A set of one million electrical traces of the SNSPD are used for PCA. (f) The rising edge time is measured using a time-to-digital converter at two threshold levels.}
    \label{fig:techniques}
\end{figure}

The first demonstration of SNSPDs showing PNR was done by Cahall et al.~\cite{cahall2017multi} in 2017. In their work, they observed a time- and photon-number dependent hotspot resistance of the nanowire. They amplify the voltage signal of the SNSPD and convert the rise time (rising edge slope or slew rate) to an amplitude with an analog inductor-resistor differentiating circuit (see Fig.~\ref{fig:techniques}(a)). Faster rise times lead to larger signal amplitudes, which are attributed to higher photon numbers. Recording a histogram of the peak height of the differentiated waveform gives rise to a sum of Gaussian distributions that can be attributed to photon number contributions of the underlying Poissonian statistics (in their case up to four photons). They identified the main limitations of PNR to be variations in the hotspot resistance and the signal-to-noise ratio. Additionally, they note that the temporal optical pulse width needs to be shorter than the stagnation time of the hotspot, otherwise no multi-photon events can be detected.

This first observation of PNR is followed up with a generalized electro-thermal model for the turn-on dynamics of SNSPDs by Nicolich et al.~\cite{nicolich2019universal}. The model (which is based on Ref.~\cite{kerman2009electrothermal}) connects the hotspot growth with features of the rising edge of the electrical readout pulse and shows a $1/\sqrt{n}$ scaling of the rising edge with photon number.

The STaND or superconducting tapered nanowire detector from Zhu et al.~\cite{zhu2020resolving} presented another approach to PNR with SNSPDs. Their detector is combined with an integrated impedance-matching taper, which makes the amplitude of the output signal of the SNSPD (see Fig.~\ref{fig:techniques}(a)) sensitive to the number of hotspots induced by absorbed photons (which was predicted by Bell et al.~\cite{bell2007photon}). The photon-number contributions in the pulse-height histograms are much more pronounced compared to Ref.~\cite{cahall2017multi}, leading to less overlap and thus better distinguishability between photon numbers up to four. Here it is again noted that impinging photons need to overlap in time (up to tens of picoseconds), as for larger optical pulse widths they observe an underestimation of the counting statistics. 
PNR has also been observed in another work using impedance-matching tapers from Colangelo et al.~\cite{colangelo2023impedance}. Additionally, Kong et al.~\cite{kong2025a} utilized an impedance-matching taper in a 16-pixel detector array design to distinguish between all possible single- and two-photon detection events of their device. The authors leverage signal-amplitude differences due to the impedance matching taper to distinguish between events where one or two photons impinge on the same pixel.

Instead of using a differentiating circuit to encode photon-number information in the amplitude of the output signal, Endo et al.~\cite{endo2021quantum} used an oscilloscope to record reference waveforms corresponding to different photon numbers, which could then be used to distinguish different events by waveform pattern matching (see Fig.~\ref{fig:techniques}(b)). As this uses a subset of the rising edge instead of a single point, they are more resistant to noise, which is especially interesting for higher photon numbers, where the overlap between waveforms increases. The authors note that new reference waveforms need to be acquired when the bias current is changed, as the bias current will change the signal amplitude and slew rate. Using their technique they can distinguish up to five photons at a wavelength of roughly $770~\mathrm{nm}$.

A similar approach to Ref.~\cite{endo2021quantum} was done by Sempere-Llagostera et al.~\cite{sempere-llagostera2022reducing}, where they use a commercial SNSPD and fit a linear function to part of the rising edge of recorded oscilloscope traces (see Fig.~\ref{fig:techniques}(c)); the extracted slopes are photon-number dependent. They use the differentiation of photon-number events to improve the second-order correlation function $g^{(2)}(0)$, by filtering multi-photon contributions from a spontaneous parametric down-conversion source. 
An alternative method to improve heralded single-photon sources was used by Davis et al.~\cite{davis2022improved}, where the photon-number information was extracted via arrival-time differences from single- and multi-photon events relative to a trigger signal (see Fig.~\ref{fig:techniques}(d)). This method enables real-time readout, and thus filtering of multi-photon events.

\begin{table}[t]
\centering
\caption{Summary of the approaches for PNR with SNSPDs. $^\dagger$The number of resolved photons as given by the authors.%; if not given, the number is taken from context. 
$^*$The aim here was the discrimination of one and more than one photon. $^\ddagger$Here the optical pulse is carved from a continuous-wave laser source.} 
\label{tab:pnr_detectors}
\begin{tabular}{p{3.75cm} p{2.1cm} p{4.0cm} p{2cm} p{1.3cm} p{1.3cm}}
\toprule
\textbf{Source} & \textbf{Detector Type} & \textbf{PNR Technique} & \textbf{Quoted PNR}$^\dagger$ & \textbf{Wave-length} & \textbf{Pulse Duration} \\
\midrule
Cahall et al.~\cite{cahall2017multi} (2017) & SNSPD & Amplitude of differentiated detection waveforms & Up to four & 1550~nm & $80$~ps$^\ddagger$ \\
\midrule
Nicolich et al.~\cite{nicolich2019universal} (2019) & SNSPD & Electrothermal model for turn-on dynamics & - & 1550~nm & $80$~ps$^\ddagger$ \\
\midrule
Zhu et al.~\cite{zhu2020resolving} (2020) & STaND & Amplitude of the output pulses & Up to five & 1550~nm & 33~ps \\
\midrule
Endo et al.~\cite{endo2021quantum} (2021) & SNSPD & Rising edge waveform pattern matching & Up to five & 772~nm & $\sim1$~ps \\
\midrule
Sempere-Llagostera et~al.~\cite{sempere-llagostera2022reducing} (2022) & Commercial SNSPD & Fitting rising edge slope & Up to two$^*$ & 1550~nm & $1$~ps \\
\midrule
Davis et al.~\cite{davis2022improved} (2022) & SNSPD & Rising edge arrival-time difference to trigger signal & Up to two$^*$ & 1550~nm & $600$~ps$^\ddagger$ \\
\midrule
Colangelo et al.~\cite{colangelo2023impedance} (2023) & Tapered SNSPD & Amplitude of the output pulses & Up to three & 775~nm, 1550~nm & 1~ps \\
\midrule
Sauer et al.~\cite{sauer2023resolving} (2023) & Commercial SNSPD & Rising and falling edge arrival-time difference to trigger signal & Up to five & 1554~nm & $\sim300$~fs \\
\midrule
Schapeler et al.~\cite{schapeler2024electrical} (2024) & Commercial SNSPD & PCA of full electrical trace; Rising and falling edge arrival-time difference to trigger signal & Up to five & 1550~nm & $\sim3$~ps \\
\midrule
Kong et al.~\cite{kong2024large} (2024) & SMSPD & Rising edge time & Up to ten & 1064~nm & $<1$~ps \\
\midrule
Los et al.~\cite{los2024high} (2024) & SNSPD & Rising edge arrival-time difference to trigger signal & Up to seven & 1064~nm & $2.3$~ps\\
\midrule
Jaha et al.~\cite{jaha2024kinetic} (2024) & Waveguide-integrated SNSPD & Rising edge arrival-time difference to trigger signal & Up to three & 1550~nm & $100$~fs\\
\midrule
Kong et al.~\cite{kong2025a} (2025) & Tapered SNSPD~array & Amplitude of the output pulses & Up to two & 1550~nm & $6$~ps\\
\midrule
Sidorova et al.~\cite{sidorova2025jitter} (2025) & Commercial SNSPD & Multi-photon jitter model & - & $1550$~nm & $1$~ps \\
\midrule
This work & Commercial SNSPD & Rising edge arrival-time difference to trigger signal & Up to three based~on~Eq.~\ref{eqn:res_fwhm} & $1550$~nm & $1$~ps \\
\bottomrule
\end{tabular}
\end{table}

Sauer et al.~\cite{sauer2023resolving} refines the retrieval of photon-number information based on arrival-time differences to a trigger signal, by utilizing both the rising and falling edge of SNSPD traces (see Fig.~\ref{fig:techniques}(d)). Two-dimensional arrival-time histograms (differences of a trigger signal to the rising edge and falling edge on either axis) reveals that a simple projection onto the rising edge does not retrieve the full photon-number information. Instead, using the additional information from the falling edge, the two-dimensional histogram can be projected on an angle, which optimizes the separation of photon-number events. This gave them discrimination up to five photons, which they used to measure the joint photon-number distribution of NOON states.

In order to find out which parts of the electrical trace from an SNSPD contains photon-number information, Schapeler et al.~\cite{schapeler2024electrical} applied principal component analysis (PCA) to a set of electrical traces recorded for different mean photon numbers per pulse (see Fig.~\ref{fig:techniques}(e)). This technique, commonly used in multivariate statistics, revealed that the photon-number information is encoded in the rising edge as well as the voltage pulse amplitude of the SNSPD trace. They identify that the photon-number information in the amplitude translates into the falling edge of the trace, which is consistent with the findings from Ref.~\cite{sauer2023resolving}. Ultimately, they conclude that a constant-threshold time-to-digital converter (time tagger) is an excellent tool to retrieve photon-number information, by measuring the arrival-time difference of the rising (and falling) edge to a trigger signal (see Fig.~\ref{fig:techniques}(d)), as this captures the relevant information with one (two) point(s) on the SNSPD trace.

Kong et al.~\cite{kong2024large} showed PNR up to ten photons, by utilizing micron-wire detectors. This is achieved by a higher inductance of the wire, which will increase the rise time of the traces. The increased rise time will lead to better distinguishability of different photon-number events, at the cost of an increased reset time of the detector. The authors implement a real-time readout of the rising edge time by splitting the electrical signal and measuring at a low and high voltage level using a time-to-digital converter (see Fig.~\ref{fig:techniques}(f)).

With the aim to investigate a wavelength range important for, e.g., quantum dots and quantum memories, Los et al.~\cite{los2024high} studies PNR with SNSPDs optimized for $850-950~\mathrm{nm}$. The authors utilize an oscilloscope to measure the distributions between rising edge of the SNSPD traces and a trigger signal (see Fig.~\ref{fig:techniques}(d)). Reducing the jitter of the SNSPD is crucial for improving PNR; as jitter reduces with increasing photon energy (lower wavelength)~\cite{korzh2020demonstration,zhang2020photon,taylor2020mid}, they achieve PNR up to seven photons at a wavelength of $1064~\mathrm{nm}$.

Waveguide-integrated SNSPDs also show PNR, as demonstrated by Jaha et al.~\cite{jaha2024kinetic}. The authors investigate PNR performance of their devices, specifically exploring the impact of kinetic inductance. They find that increasing the kinetic inductance of the detector improves PNR, as it increases the distance between photon-number peaks in arrival-time measurements relative to a trigger signal (see Fig.~\ref{fig:techniques}(d)). Again, this improvement of PNR comes at the cost of an increased reset time, leading to a reduced maximum count rate of the detector, which is consistent with findings from Kong et al.~\cite{kong2024large}.

Sidorova et al.~\cite{sidorova2025jitter} formulates a quantitative physical model of the multi-photon absorption in SNSPDs. They discuss the detection mechanism, hotspot formation, intrinsic jitter and dynamics of resistive domains for simultaneous and delayed multi-photon absorption in the same site as well as independent sites. Their model is based on a sum of exponentially-modified Gaussian distributions and merely three fitting parameters. Applying their model to experimentally measured arrival-time histograms (see Fig.~\ref{fig:techniques}(d)) from a commercial SNSPD, they find excellent agreement over orders of magnitude in probability and mean photon number. 

\section{Photon-number resolvability}
For the papers mentioned in Sec.~\ref{sec:stateoftheart}, we quote the number of resolved photons as stated by the authors (see Table~\ref{tab:pnr_detectors}). However, there exists no accepted definition of this number, which makes comparisons difficult. In what follows, we propose a resolvability criterion which can be readily applied to PNR detectors based on measuring continuous distributions dependent on photon number.

We define the maximum resolvability of an intrinsic PNR detector to be the highest photon number $n$ which has a peak with a full width at half maximum (FWHM) that is smaller than the difference of the peak centers $\mu$ for photon number contributions of $n$ and $(n+1)$. Thus the maximal photon number the detector may resolve is the photon number $n$ for which the inequality
\begin{equation}\label{eqn:res_fwhm}
    (\mu_n-\mu_{n+1}) \geq \mathrm{FWHM}_n
\end{equation}
is still valid. This measure is independent of the photon-number retrieval method (see Sec.~\ref{sec:stateoftheart}), i.e., whether the differentiation of photon numbers is done over the time axis or voltage axis.

Arrival-time histograms of the rising edge of SNSPDs are well described by a sum of exponentially-modified Gaussian (EMG) distributions~\cite{sidorova2025jitter}. The total width $\sigma_{\text{tot},n}=\sqrt{\sigma_n^2+\tau_n^2}$ of the underlying distribution for a specific photon-number component $n$ is determined by the Gaussian $\sigma_n$ and exponential $\tau_n$ components (both of which may be photon-number dependent). 
The Gaussian component itself consists of jitter contributions arising from noise, instrumentation, optical pulse duration, detector geometry and intrinsic jitter
\begin{equation}\label{eqn:sigma}
\sigma_n = \sqrt{\sigma_{\text{noise}}^2 + \sigma_{\text{inst}}^2+\sigma_\textrm{opt}^2 + \sigma_{\text{geom},n}^2 + \sigma_{\textrm{int},n}^2}\,,
\end{equation}
of which the geometric jitter and the intrinsic jitter have photon-number dependence~\cite{sidorova2025jitter}.

Due to the exponential tail of the EMG distribution, there is no direct connection between the standard deviation of the distribution ($\sigma_{\text{tot},n}$) and the FWHM. Only if the exponential component is sufficiently small ($\tau_n\ll\sigma_n$) one can use the standard deviation to solve the inequality
\begin{align}\label{eqn:res_gauss}
   (\mu_n-\mu_{n+1}) &\geq k\sigma_{\mathrm{tot},n}    \\
     \left(\frac{\Delta\mu}{\sqrt{n}}-\frac{\Delta\mu}{\sqrt{n+1}}\right) &\geq k\sqrt{\sigma_n^2+\tau_n^2}\,,
\end{align}
where $k=2\sqrt{2\mathrm{ln}(2)}\approx2.355$ for the FWHM of a Gaussian distribution and we used that the separation of photon-number peaks in the response of an SNSPD follows a $1/\sqrt{n}$ dependence~\cite{nicolich2019universal}.

\begin{figure}
    \centering
    \includegraphics[width=1\linewidth]{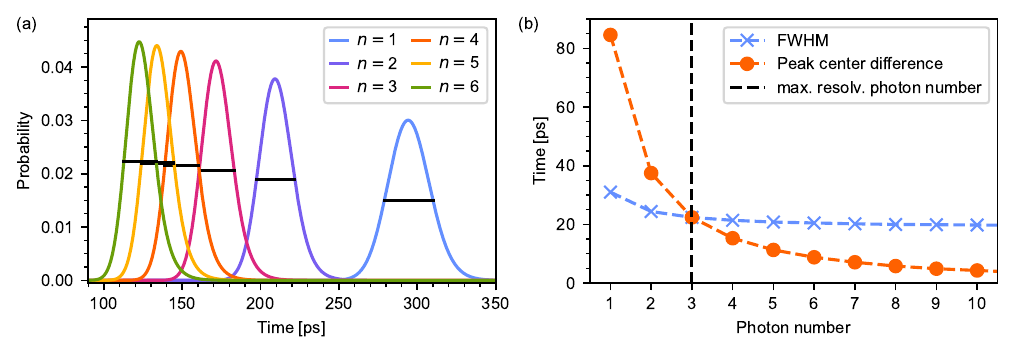}
    \caption{(a) Exponentially-modified Gaussian distributions for photon-number contributions from one to six (based on the parameters from Ref.~\cite{sidorova2025jitter} and normalized to one) with their FWHM indicated by the black solid line. (b) Peak center difference ($\mu_n-\mu_{n+1}$) and $\mathrm{FWHM}_n$ as a function of photon number $n$. The inequality from Eq.~\ref{eqn:res_fwhm} is valid until $n=3$ (marked by the black dashed vertical line).}
    \label{fig:fwhm_plot}
\end{figure}

By measuring the jitter components from Eq.~\ref{eqn:sigma} and applying the model from Ref.~\cite{sidorova2025jitter} to experimental data to find the fitting parameters $\sigma_\mathrm{int}$, $\tau$ and $\Delta\mu$ one can directly determine the photon-number-resolution limit of a detector. Taking all parameters from Ref.~\cite{sidorova2025jitter} (which are based on our experimental data) we show the EMG distributions (normalized to one) for the first six photon-number contributions (see Fig.~\ref{fig:fwhm_plot}(a)). The distributions for three and four photons intersect approximately at their FWHM, indicating that this is the resolution limit. By applying Eq.~\ref{eqn:res_fwhm}, i.e., investigating the peak center difference and the FWHM as a function of $n$ (see Fig.~\ref{fig:fwhm_plot}(b)), we find that the SNSPD can resolve up to $n=3$ photons and every event that contributes to earlier times in the arrival-time histogram may be regarded as a ``four or more photon'' event. 

\section{Outlook}
From the different techniques presented in Sec.~\ref{sec:stateoftheart}, it is clear that there are many ways to retrieve the intrinsic photon-number information encoded in the electrical traces of SNSPDs. Certain techniques might be more suitable for specific tasks, e.g., whether a trigger signal can be used, whether the information needs to be read-out in real time, or how much differentiation between photon numbers is necessary for a certain experiment. One pertinent example is heralding single-photons from spontaneous parametric down-conversion (SPDC) or spontaneous four-wave mixing (SFWM). To minimize the effect of heralding on higher-order terms, it is sufficient to distinguish between one or more than one photon. Therefore, in order to minimize the assignment error, the separation of these events needs to be large. Alternatively, for photon-counting tasks, one might accept slightly less differentiation between photon-number events.

The most resolved photons and the best separation between photon-number events were shown by Kong et al.~\cite{kong2024large} using a micron-wire detector with a large kinetic inductance. However, this improvement in PNR came at the cost of reduced repetition rate due to a slower reset time of the detector. For many applications, it is desirable to have high repetition rates.
Although, at high count rates, a time-walk effect~\cite{mueller2022time} (arising from count rate dependent variations in the bias current) can occur when the inter-arrival time of optical pulses is in the order of the reset time of the device. This leads to distorted electrical traces and can limit the PNR capability of a detector. This is because all techniques to retrieve intrinsic PNR (see Fig.~\ref{fig:techniques}) rely on the assumption that the detector starts from the same initial condition from shot to shot; additional uncertainty will lead to less distinguishable photon-number events.

Another challenge is to extend the number of resolvable photons beyond ten. This may be done by combining multiplexing schemes (spatially or temporally) with SNSPDs with intrinsic PNR. The most promising way would likely be to use integrated SNSPD arrays, where each pixel can resolve a limited number of photons. Reading out every pixel separately will then increase the resolution up to larger photon numbers. This however comes at the cost of increased numbers of readout lines, which is not always compatible with the limited heat-load of cryostats. Therefore, the optimal solution would likely be a combination of the photon-number events from individual pixels at cryogenic temperatures through some form of logic, and then reading out this information through a single readout line.

Lastly, it is desirable that PNR works independent of the wavelength of the photons (which has already been shown by optimizing the SNSPD for specific wavelengths, as can be seen in Table~\ref{tab:pnr_detectors}) and that the resolution of photon number using SNSPDs can be extended to pulses of longer duration, which depends on the superconducting material and substrate properties.

In this paper we have reviewed the state-of-the-art of intrinsic PNR with SNSPDs and provided an overview of the experimental techniques to retrieve the photon-number information (see Fig.~\ref{fig:techniques} and Table~\ref{tab:pnr_detectors}). We have then proposed a resolvability criterion to PNR based on SNSPDs, to make comparison between literature more straightforward. Finally, we gave an outlook on remaining challenges for photon-number resolution based on SNSPDs.

\acknowledgments
Partially funded by the European Union (ERC, QuESADILLA, 101042399). Views and opinions expressed are however those of the author(s) only and do not necessarily reflect those of the European Union or the European Research Council Executive Agency. Neither the European Union nor the granting authority can be held responsible for them. This work has received funding from the German Ministry of Education and Research within the PhoQuant project (grant number 13N16103). F.S. is part of the Max Planck School of Photonics supported by the German Federal Ministry of Education and Research (BMBF), the Max Planck Society, and the Fraunhofer Society.

% References
\bibliography{report} % bibliography data in report.bib

\begin{thebibliography}{10}

\bibitem{gisin2002quantum}
Gisin, N., Ribordy, G., Tittel, W., and Zbinden, H., ``{Quantum cryptography},'' {\em Reviews of Modern Physics}~{\bf 74}(1),  145--195 (2002).

\bibitem{gisin2007quantum}
Gisin, N. and Thew, R., ``{Quantum communication},'' {\em Nature Photonics}~{\bf 1}(3),  165--171 (2007).

\bibitem{kok2007linear}
Kok, P., Munro, W.~J., Nemoto, K., Ralph, T.~C., Dowling, J.~P., and Milburn, G.~J., ``{Linear optical quantum computing with photonic qubits},'' {\em Reviews of Modern Physics}~{\bf 79}(1),  135--174 (2007).

\bibitem{slussarenko2017unconditional}
Slussarenko, S., Weston, M.~M., Chrzanowski, H.~M., Shalm, L.~K., Verma, V.~B., Nam, S.~W., and Pryde, G.~J., ``{Unconditional violation of the shot-noise limit in photonic quantum metrology},'' {\em Nature Photonics}~{\bf 11}(11),  700--703 (2017).

\bibitem{cabrera1998detection}
Cabrera, B., Clarke, R.~M., Colling, P., Miller, A.~J., Nam, S., and Romani, R.~W., ``{Detection of single infrared, optical, and ultraviolet photons using superconducting transition edge sensors},'' {\em Applied Physics Letters}~{\bf 73}(6),  735--737 (1998).

\bibitem{lita2008counting}
Lita, A.~E., Miller, A.~J., and Nam, S.~W., ``{Counting near-infrared single-photons with 95\% efficiency},'' {\em Optics Express}~{\bf 16}(5),  3032 (2008).

\bibitem{reddy2020superconducting}
Reddy, D.~V., Nerem, R.~R., Nam, S.~W., Mirin, R.~P., and Verma, V.~B., ``{Superconducting nanowire single-photon detectors with 98\% system detection efficiency at 1550 nm},'' {\em Optica}~{\bf 7}(12),  1649 (2020).

\bibitem{chang2021detecting}
Chang, J., Los, J. W.~N., Tenorio-Pearl, J.~O., Noordzij, N., Gourgues, R., Guardiani, A., Zichi, J.~R., Pereira, S.~F., Urbach, H.~P., Zwiller, V., Dorenbos, S.~N., and {Esmaeil Zadeh}, I., ``{Detecting telecom single photons with 99.5-2.07+0.5\% system detection efficiency and high time resolution},'' {\em APL Photonics}~{\bf 6}(3),  36114 (2021).

\bibitem{korzh2020demonstration}
Korzh, B., Zhao, Q.-Y., Allmaras, J.~P., Frasca, S., Autry, T.~M., Bersin, E.~A., Beyer, A.~D., Briggs, R.~M., Bumble, B., Colangelo, M., Crouch, G.~M., Dane, A.~E., Gerrits, T., Lita, A.~E., Marsili, F., Moody, G., Pe{\~{n}}a, C., Ramirez, E., Rezac, J.~D., Sinclair, N., Stevens, M.~J., Velasco, A.~E., Verma, V.~B., Wollman, E.~E., Xie, S., Zhu, D., Hale, P.~D., Spiropulu, M., Silverman, K.~L., Mirin, R.~P., Nam, S.~W., Kozorezov, A.~G., Shaw, M.~D., and Berggren, K.~K., ``{Demonstration of sub-3 ps temporal resolution with a superconducting nanowire single-photon detector},'' {\em Nature Photonics}~{\bf 14}(4),  250--255 (2020).

\bibitem{muenzberg2018superconducting}
M{\"{u}}nzberg, J., Vetter, A., Beutel, F., Hartmann, W., Ferrari, S., Pernice, W. H.~P., and Rockstuhl, C., ``{Superconducting nanowire single-photon detector implemented in a 2D photonic crystal cavity},'' {\em Optica}~{\bf 5}(5),  658--665 (2018).

\bibitem{chiles2022new}
Chiles, J., Charaev, I., Lasenby, R., Baryakhtar, M., Huang, J., Roshko, A., Burton, G., Colangelo, M., {Van Tilburg}, K., Arvanitaki, A., Nam, S.~W., and Berggren, K.~K., ``{New Constraints on Dark Photon Dark Matter with Superconducting Nanowire Detectors in an Optical Haloscope},'' {\em Physical Review Letters}~{\bf 128}(23),  231802 (2022).

\bibitem{cahall2017multi}
Cahall, C., Nicolich, K.~L., Islam, N.~T., Lafyatis, G.~P., Miller, A.~J., Gauthier, D.~J., and Kim, J., ``{Multi-photon detection using a conventional superconducting nanowire single-photon detector},'' {\em Optica}~{\bf 4}(12),  1534--1535 (2017).

\bibitem{nicolich2019universal}
Nicolich, K.~L., Cahall, C., Islam, N.~T., Lafyatis, G.~P., Kim, J., Miller, A.~J., and Gauthier, D.~J., ``{Universal Model for the Turn-On Dynamics of Superconducting Nanowire Single-Photon Detectors},'' {\em Physical Review Applied}~{\bf 12}(3),  34020 (2019).

\bibitem{kerman2009electrothermal}
Kerman, A.~J., Yang, J. K.~W., Molnar, R.~J., Dauler, E.~A., and Berggren, K.~K., ``{Electrothermal feedback in superconducting nanowire single-photon detectors},'' {\em Physical Review B}~{\bf 79}(10),  100509 (2009).

\bibitem{zhu2020resolving}
Zhu, D., Colangelo, M., Chen, C., Korzh, B.~A., Wong, F. N.~C., Shaw, M.~D., and Berggren, K.~K., ``{Resolving Photon Numbers Using a Superconducting Nanowire with Impedance-Matching Taper},'' {\em Nano Letters}~{\bf 20}(5),  3858--3863 (2020).

\bibitem{bell2007photon}
Bell, M., Antipov, A., Karasik, B., Sergeev, A., Mitin, V., and Verevkin, A., ``{Photon Number-Resolved Detection With Sequentially Connected Nanowires},'' {\em IEEE Transactions on Applied Superconductivity}~{\bf 17}(2),  289--292 (2007).

\bibitem{colangelo2023impedance}
Colangelo, M., Korzh, B., Allmaras, J.~P., Beyer, A.~D., Mueller, A.~S., Briggs, R.~M., Bumble, B., Runyan, M., Stevens, M.~J., McCaughan, A.~N., Zhu, D., Smith, S., Becker, W., Narv{\'{a}}ez, L., Bienfang, J.~C., Frasca, S., Velasco, A.~E., Ramirez, E.~E., Walter, A.~B., Schmidt, E., Wollman, E.~E., Spiropulu, M., Mirin, R., Nam, S.~W., Berggren, K.~K., and Shaw, M.~D., ``{Impedance-Matched Differential Superconducting Nanowire Detectors},'' {\em Physical Review Applied}~{\bf 19}(4),  044093 (2023).

\bibitem{kong2025a}
Kong, L.-D., Zhang, T.-Z., Liu, X.-Y., Zhao, X., Xiong, J.-M., Li, H., Wang, Z., Xie, X.-M., and You, L.-X., ``{A superconducting nanowire two-photon coincidence counter with combinatorial time logic and amplitude multiplexing},'' {\em Nature Photonics}  (2025).

\bibitem{endo2021quantum}
Endo, M., Sonoyama, T., Matsuyama, M., Okamoto, F., Miki, S., Yabuno, M., China, F., Terai, H., and Furusawa, A., ``{Quantum detector tomography of a superconducting nanostrip photon-number-resolving detector},'' {\em Optics Express}~{\bf 29}(8),  11728 (2021).

\bibitem{sempere-llagostera2022reducing}
Sempere-Llagostera, S., Thekkadath, G.~S., Patel, R.~B., Kolthammer, W.~S., and Walmsley, I.~A., ``{Reducing g(2)(0) of a parametric down-conversion source via photon-number resolution with superconducting nanowire detectors},'' {\em Optics Express}~{\bf 30}(2),  3138--3147 (2022).

\bibitem{davis2022improved}
Davis, S.~I., Mueller, A., Valivarthi, R., Lauk, N., Narvaez, L., Korzh, B., Beyer, A.~D., Cerri, O., Colangelo, M., Berggren, K.~K., Shaw, M.~D., Xie, S., Sinclair, N., and Spiropulu, M., ``{Improved Heralded Single-Photon Source with a Photon-Number-Resolving Superconducting Nanowire Detector},'' {\em Physical Review Applied}~{\bf 18}(6),  64007 (2022).

\bibitem{sauer2023resolving}
Sauer, G., Kolarczik, M., Gomez, R., Conrad, J., and Steinlechner, F., ``{Resolving Photon Numbers Using Ultra-High-Resolution Timing of a Single Low-Jitter Superconducting Nanowire Detector},'' {\em arXiv preprint arXiv:2310.12472}  (2023).

\bibitem{schapeler2024electrical}
Schapeler, T., Lamberty, N., Hummel, T., Schlue, F., Stefszky, M., Brecht, B., Silberhorn, C., and Bartley, T.~J., ``{Electrical trace analysis of superconducting nanowire photon-number-resolving detectors},'' {\em Physical Review Applied}~{\bf 22}(1),  14024 (2024).

\bibitem{kong2024large}
Kong, L.-D., Zhang, T.-Z., Liu, X.-Y., Li, H., Wang, Z., Xie, X.-M., and You, L.-X., ``{Large-inductance superconducting microstrip photon detector enabling 10 photon-number resolution},'' {\em Advanced Photonics}~{\bf 6}(1),  16004 (2024).

\bibitem{los2024high}
Los, J. W.~N., Sidorova, M., Lopez-Rodriguez, B., Qualm, P., Chang, J., Steinhauer, S., Zwiller, V., and Zadeh, I.~E., ``{High-performance photon number resolving detectors for 850–950 nm wavelength range},'' {\em APL Photonics}~{\bf 9}(6) (2024).

\bibitem{jaha2024kinetic}
Jaha, R., Graham-Scott, C.~A., Abazi, A.~S., Pernice, W., Schuck, C., and Ferrari, S., ``{Kinetic Inductance and Jitter Dependence of the Intrinsic Photon Number Resolution in Superconducting Nanowire Single-Photon Detectors},'' {\em arXiv preprint arXiv:2410.23162}  (2024).

\bibitem{sidorova2025jitter}
Sidorova, M., Schapeler, T., Semenov, A.~D., Schlue, F., Stefszky, M., Brecht, B., Silberhorn, C., and Bartley, T.~J., ``Jitter in photon-number-resolved detection by superconducting nanowires,'' {\em arXiv quant-ph/2503.17146}  (2025).

\bibitem{zhang2020photon}
Zhang, H., Liu, J., Guo, J., Xiao, L., and Xie, J., ``{Photon energy-dependent timing jitter and spectrum resolution research based on time-resolved SNSPDs},'' {\em Optics Express}~{\bf 28}(11),  16696--16707 (2020).

\bibitem{taylor2020mid}
Taylor, G.~G., MacKenzie, E.~N., Korzh, B., Morozov, D.~V., Bumble, B., Beyer, A.~D., Allmaras, J.~P., Shaw, M.~D., and Hadfield, R.~H., ``{Mid-infrared timing jitter of superconducting nanowire single-photon detectors},'' {\em Applied Physics Letters}~{\bf 121}(21),  214001 (2022).

\bibitem{mueller2022time}
Mueller, A., Wollman, E.~E., Korzh, B., Beyer, A.~D., Narvaez, L., Rogalin, R., Spiropulu, M., and Shaw, M.~D., ``{Time-walk and jitter correction in SNSPDs at high count rates},'' {\em Applied Physics Letters}~{\bf 122}(4) (2023).

\end{thebibliography}
\bibliographystyle{spiebib} % makes bibtex use spiebib.bst

\end{document}